\documentstyle[aps,pre,psfig]{revtex}

\begin{document} 

%\draft

\renewcommand{\baselinestretch}{1.0}
%\twocolumn
%\normalsize

\title{The Control of high-dimensional Chaos in Time-Delay Systems} 
 
\author{{\em M. J. B\"unner}	\thanks{present address: Istituto Nazionale di Ottica, 
	Largo E. Fermi 6, I-50125 Firenze,
	Italy,e-mail: buenner@fox.ino.it} \\
	{\em Max-Planck-Institut f\"ur Physik komplexer Systeme,} \\ 
	{\em N\"othnitzer Str. 38,} \\ 
	{\em D-01187 Dresden, Germany.} \\ 
	} 
 
\date{Septembre, 1998} 
 
\maketitle 
 
\begin{abstract} 

We present the control of the high-dimensional chaos, with possibly a 
large number of positive Lyapunov-exponents, of unknown time-delay systems 
to an arbitrary goal dynamics. 
We give  an  existence-and-uniqueness theorem for the control force.
In the case of an unknown system, a formula to compute a  model-based control force
is derived.
    % a global model obtained with the help of a previously published 
    %identification procedure. The stability of the
    %control with respect to an imperfect modeling, as well as noise, is discussed.  
    %%Additionally, we achieve the control of unstable periodic orbits with
    %the help of a minimization.
We give an example by demonstrating the control of the Mackey-Glass system
towards a fixed point and a R\"ossler-dynamics.

\end{abstract} 
\pacs{PACS numbers: 02.30.Ks, 05.45.+b, 07.05.Dz}

{\bf We treat the control problem for nonlinear time-delay systems. The aim of the control
is to bring the system  from a given dynamical state to
an arbitrary goal dynamics. 
We give the conditions under which the control force uniquely exists. 
In the case of an unknown nonlinear delay system, we introduce a procedure to compute
the control force with the help of a previously published 
identification procedure.
The results are also applicable in the case of the very high-dimensional chaotic
motion, with possibly a large number of positive Lyapunov exponents,  
typically observed in nonlinear delay systems. 
Stability towards noise and towards an imperfect modeling is treated in linear response.
Additionally, we discuss the control of unstable periodic orbits with
the help of a minimization. A numerical example, the control of the 
Mackey-Glass system from a very high-dimensional chaotic state  towards
a stationary state or a non-periodic time-varying state 
(the low-dimensional chaotic motion observed in the R\"ossler system), is presented 
in detail. }

%We present the control of the high-dimensional chaos, with possibly a 
%large number of positive Lyapunov-exponents, of unknown time-delay systems 
%to an arbitrary goal dynamics. 
%We discuss the  existence and uniqueness of the control force.
%In the case of an unknown system, the control relies on
% a global model obtained with the help of a previously published 
%identification procedure. The stability of the
%control with respect to an imperfect modeling, as well as noise, is discussed.  
%Additionally, we achieve the control of unstable periodic orbits with
%the help of a minimization.
%We give an example by demonstrating the control of the Mackey-Glass system
%towards a fixed point and to the low-dimensional chaotic motion observed
%in the R\"ossler system.

\section{Introduction}

The possibility of controlling unstable periodic orbits (UPOs) of chaotic 
systems with the help of suitably
chosen external perturbances, which vanish in the case of a successful 
control, initiated a vast research activity.
The OGY-method \cite{OGY} and the Pyragas-method \cite{Pyragas} 
attracted most attention, so far, and 
have been successfully applied to various experimental situations.
While  the OGY-method applies a local modeling, 
% and is, in general, restricted to the control of low-dimensional chaos, 
the Pyragas-method does not apply any 
modeling and is restricted by the value of the imaginary
part of the Floquet-exponents of the unstable periodic orbit to be controlled
\cite{Just97}. 
One of the aims of the control of UPOs has been to enable the switching
between different states with the help of small perturbations only. In low-dimensional
chaotic states (with only one positive Lyapunov exponent), 
though, the basic frequencies of most of the UPOs 
are nearly identical and a lot of UPOs are therefore 
quite similar.  It is the advantage of hyperchaotic chaotic states 
(with more than one positive Lyapunov exponent) to
provide for a large variety of UPOs with different basic frequencies.
For this reason, the control of UPOs of hyperchaotic chaos is desirable.
The successful control of UPOs of high-dimensional chaos with either a 
single control force \cite{ses-single},
or a multitude of control forces  \cite{ses-multiple} 
has been reported. More specifically, the control of of UPOs of time-delay systems also
in hyperchaotic regimes has been achieved with a single control force only 
\cite{Pyragas,tds-control}. Despite that remarkable success in specific cases, it remains
an open problem if all hyperchaotic states  can be controlled by a single control force, and
what actually  determines the number of necessary control variables. 
In other words - under which {\em necessary} and {\em sufficient} conditions does
a (scalar or multi-variate) force, which controls a hyperchaotic state towards 
a UPO exist?
For the latter type of control we use the term 'control of UPOs' throughout this
paper.

On the other hand, we use the term 'control' to describe the more general task to
force a system to an arbitrary, predefined
goal dynamics, where the control force is not vanishing. The control to
a predefined goal dynamics is also the more general, as well as the practically
more relevant case, since most technical processes, such as manufacturing, chemical
engineering, etc. rely on this type of control.
It has been treated in engineering and applied mathematics
for a long time. While for the control of linear systems
a sufficient understanding has been achieved, the control of nonlinear systems,
in general, remains an open problem \cite{control}. 
More specifically, the control of linear time-delay systems were summarized by
Marshall \cite{Marshall75}.
Along these lines, the control of chaotic states in ordinary differential
equations has been demonstrated \cite{H"ubler}. 
In the case of an unknown time-evolution equation, though, the method 
has to rely on a modeling, i.e. with the help of Takens-like techniques, 
and is therefore restricted to low-dimensional chaotic states in typical situations.
The purpose of this paper is to show that the very high-dimensional 
chaotic states \cite{high-dim-chaos} 
of a possibly unknown time-delay system with $N$ components can be
controlled to an arbitrary goal dynamics, with $N$ control forces, for the following
reasons: (1) a global model of an unknown,
$N$-component time-delay system can be obtained with the help of $N$ time series
only, as shown in \cite{BuennerPRE96}-\cite{Voss97}.
(2) The global model allows to compensate the instability-inducing effects of
the time delay by suitably chosen $N$ control forces as will be shown in this paper.
The paper is organized as follows:
At first we proove under which conditions a unique control force exists, which
allows for a stable control of a time-delay system.  
This result is then used to derive a control scheme for the
control of a time-delay 
system with {\em unknown} equations where a global model
is achieved with an adequate identification 
procedure. Finally, we show the result of
a computer experiment.

\section{The control of unknown Delay Systems}

Suppose the dynamics of the system to be controlled is determined by a
$N$-component, non-autonomous, time-delayed differential equation 
\begin{equation}
\label{tddekont}
\dot{y}=H(y,y_{\tau_0},F(t)), \hspace{2.0cm} y_{\tau_0}=y(t-\tau_0), 
\end{equation}
where $y(t) \in {\cal D}_1 \subset {\cal R}^N$ is the range of the dynamics,  
$F(t) \in {\cal D}_2 \subset {\cal R}^N$
is the range of the external force. 
Suppose the function $H: {\cal R}^{3N} \rightarrow {\cal R}^N$ is a continuous  
and invertible in ${\cal D}_2$ with respect to the third argument.
The control problem consists in proving the existence of a force $F(t)$, which makes a predefined goal
dynamics $z(t)$ a stable solution of eq. (\ref{tddekont}). In practice, global 
stability is desirable, since that excludes coexisting dynamical states.
This can be achieved by imposing an
additional condition on the dynamics $y(t)$ in the form of a 
first-order, ordinary, non-autonomous differential equation 
\begin{equation}
\label{g}
\dot{y}=g[z(t)](y), \hspace{2.0cm} y(0)=y_0,
\end{equation}
with  $g[z](y) \in {\cal D}_2$ for $y(t) \in {\cal D}_1$. 
The function $g: {\cal R}^{N} \rightarrow {\cal R}^N$ has a functional dependence on the goal dynamics $z(t)$. 
We require $z(t)$ to be a globally stable solution of eq. (\ref{g}): 
$\dot{z}=g[z](z)$, with $y \in {\cal D}_1 \rightarrow z$, for $t \rightarrow \infty$.
To ensure the stability of eq. (2) towards  a small amount of noise (linear response regime),
we also require local stability:
$\Re(\lambda_i(z(t))) < 0, \forall t, i=1,...,N$, where $\lambda_i(z(t))$ are the
time-dependent eigenvalues of the time-dependent Jacobian of system (\ref{g}). 
%$\frac{\partial}{\partial y} g[z](y=z(t))} < 0, \forall t$. 
Equating the two requirements upon the dynamics we get
a condition for the control force $F(t)$,
\begin{equation}
\label{forcecond}
g[z](y)=H(y,y_{\tau_0},F(t)).
\end{equation}
%If the function $H$ is invertible in ${\cal D}_2$ with respect to the third
%argument for $(y,y_{\tau_0}) \in {\cal D}_1$
Since we require invertibility of the function $H$, the control force $F(t)$ uniquely
exists for all values of the delay time $\tau_0$ and can be determined via
\begin{equation}
\label{force}
F(t)=H^{-1}(y,y_{\tau_0},g[z](y)).
\end{equation}
The control force cancels the instability-inducing effects of the time-delay. 
The dynamics is only determined by eq. (\ref{g}), and inherits its
stability properties. Therefore, the goal dynamics $z(t)$ is globally stable, and stable
in linear response to a small amount of additional noise.  
%The control force $F$, leading to a stable control, exists uniquely for all 
%values of the delay time $\tau_0$. 
But, the stability range might decrease for an increasing
value of $\tau_0$, leading to an increased sensitivity towards
noise for an increasing $\tau_0$.
%Also, the
%stability towards a small amount of noise (linear response regime) 
%can be ensured by a proper choice of $g$.

From the practical point of view it is essential to discuss the control of
delay systems, whose time-delay differential equations  are not known, but whose 
dynamics $y(t)$ is accessible to measurement and to an external driving via
the control force $F(t)$. 
For the ease of presentation, we focus on scalar time-delay systems with a single,
discrete delay time $\tau_0$. We emphasize, though, that the presented ideas
are also applicable in the case of multi-component time-delay systems, or
systems with multiple delay times or a distribution of delay times.
The central task is to find 
a model equation for the dynamics: $F(t)=H_r^{-1}(y,y_{\tau_r},\dot{y})$, where the
function $H_r^{-1}$ and the delay 
time $\tau_r$ are unknown and have to be determined
from the dynamics. It has been shown, recently, that a global model of 
autonomous scalar time-delay
systems can be obtained by measuring  a scalar component only, even if the dynamics is
very high-dimensional chaotic \cite{BuennerPRE96}-\cite{Voss97}.
The method relies on the detection of nonlinear correlations
of the observable, its time derivative and its time-delayed value.
For the problem under consideration, 
a slight modification of this procedure is appropriate. 
The basic idea is to disturb the system via a time-dependent external force
$F(t) \in {\cal D}_F$ and measure the systems reaction 
$y(t) \in {\cal D}_1, \dot{y}(t) \in {\cal D}'_1$. 
Then, one has to find the delay time $\tau_r$, for which nonlinear
correlations in the form $F(t)=H_r^{-1}(y,y_{\tau},\dot{y})$ exist.
Adequate measures to detect 
nonlinear correlations are ,i.e., the filling factor 
\cite{BuennerPRE96,BuennerPRE97}, the 
maximal correlation \cite{Voss97}, or the forecast error \cite{Hegger98}.
The function 
$H_r^{-1}$ can be determined with the help of adequate fitting procedures.
The control force is then given by
\begin{equation}
\label{force_r}
F(t)=H^{-1}_r(y,y_{\tau_0},g[z](y)),
\end{equation}
with $(y,y_{\tau}) \in {\cal D}_1$, and $g[z](y) \in  {\cal D}'_1$ if $H_r$ is invertible.
There are an infinite number of functions $g$, which guarantee the existence of the control force. 
In practice, $g$ can be chosen to meet the purpose best, a very simple choice will be:
$g[z](y)=-1/T (y-(z+T\dot{z}))$, with a single  control parameter $T$. 
If the function $H_r$ is not invertible in the required range,  
there exists no control force leading to a stable control. 
with the chosen $g$.
Possibly, control can be achieved, then,  by changing $g$ or 
the action of the control upon the dynamics (via $H_r^{-1}$). 
If the function $H_r$ is
locally non-unique invertible in the required range, the control force exists
in several regimes and we can define a piecewise
continuous control force $F(t)$ similar to eq. (\ref{force_r}). In this case  special care
has to be taken for the values $(y,y_{\tau_0})$ for which the derivative of $H_r$
with respect to the third argument vanishes. 

Since, in general, the model
$H_r^{-1}$ slightly deviates from the actual dynamics
$H^{-1}=H^{-1}_r - \epsilon \bar{H}^{-1}$, with $\epsilon$ being small, 
it is crucial to discuss the stability of the control towards 
the deviation $\epsilon \bar{H}^{-1}$.
For the ease
of presentation, we restrict our discussion to the case of an additive action
of the control force.
Then, the dynamics of the system with control is determined by the
non-autonomuous time-delay equation $\dot{y}=  \epsilon
\bar{h}(y,y_{\tau})+g[z](y)$. Calculating the deviations $\Delta(t)$, with
$y  \equiv z + \epsilon  \Delta$ in linear response, we end up with the differential
equation: $\dot{\Delta}=\frac{\partial g}{\partial y}(z,z) \Delta +
\bar{h}(y,y_{\tau})$. As expected, the quality of the control is directly related to
the quality of the model. Deviations from the goal dynamics are proportional
to deviations of  the model as long as the linear-response approximations are valid. If 
not, the deviations of the model might lead to a failure of
the control. Since we require the goal dynamics $z(t)$ to be a globally stable
solution of (\ref{g}), and $H_r^{-1}$ is a global model of the delayed system,
the goal dynamics will, in typical cases, 
also be a globally stable solution of (\ref{tddekont}) with
the control force (\ref{force_r}).
This is a huge advantage compared to a control, which only
uses  a local modeling, where the
control might fail because of coexisting attractors; a situation typically encountered
in high-dimensional chaotic systems.

\section{The Control of the Mackey-Glass System}

In order to demonstrate the control, we apply the scheme to 
the Mackey-Glass system, $\dot{y}=f(y_{\tau_0})-y+F(t), 
f(y_{\tau_0})=\frac{3y}{1+y_{\tau_0}^{10}}$, 
in a computer experiment 
for several values of the delay time $\tau_0$ and several goal dynamics $z(t)$. 
The model has been proposed to account for 
the observed large amplitude oscillations of 
the number of circulating white blood cells of patients suffering from chronic
granulocyctic leukemia (first Ref. of \cite{physiology}). 
We assume the action of the control force $F(t)$ to be additive and to be known.
Therefore, it is sufficient to 
model the system from a time series of the uncontrolled system and subsequently
apply the model to compute the control force $F(t)$ via eq. (\ref{force_r}).
Because of the demonstrational character of the example we skip most of the details
of the identification in order to keep the presentation short.
At first, we computed a trajectory of the uncontrolled system ($F=0$) for $\tau_0=80.00$ 
with a Runge-Kutta algorithm of
fourth order (transient time: $100*\tau_0$; length of the trajectory: $100*\tau_0$ with
10.000 data points). The range has been 
${\cal D}_1=[0.20; 2.02]$. The uncontrolled system
exhibits a high-dimensional chaotic state. 
To estimate the delay time we used a filling factor analysis 
\cite{BuennerPRE96}. The $\tau$-dependent filling factor is shown in Fig. 1(a), 
where the delay time is indicated by a sharp
local minimum. The data allow for a non-erronous estimate 
$\tau_r=80.00 \pm 0.01$. In the next step, the function
$h_r(y,y_{\tau_r})=f_r(y_{\tau_r})-y$ has to be determined. 
For the fit of the function $f_r$
we used a rational function with a polynomial of sixth 
order in the nominator and a polynomial of second order
in the denominator. The quadratic error has been $1.59*10^{-4}$. 
In Fig. 1(b) the difference of the two functions 
$\bar{f}=f-f_r$ is shown. 

We choose the requirement upon the dynamics 
(see eq. (\ref{g})), $g$,  in a simple linear form, 
\begin{equation}
\label{g_linear}
\dot{y}=g[z](y)=-1/T (y-(z+T\dot{z})), 
\end{equation}
with a single control parameter $T$. The goal 
dynamics $z(t)$ is a globally stable solution
of (\ref{g_linear}) for $y \in {\cal R}$.
Therefore, we arrive at a non-autonomous time-delay 
equation for the controlled system
\begin{equation}
\label{mkg1}
\dot{y}=\bar{f}(y_{\tau_0})- \frac{1}{T}(y-(z+T\dot{z})).
\end{equation}
At first we wish to discuss the control of 
a fixed point, $z(t)=z_0, \dot{z}=0$, since in this
case the action of the control is most obvious. 
In Fig. 2 we show the control of a 
chaotic state ($\tau=80.0$) to the fixed point 
$z_0=1.5$. Note that for $\tau=80.0$ the dynamical state of the uncontrolled system
is very high-dimensional with 49 positive Lyapunov-exponents and a Kaplan-Yorke
Dimension of 86.  
The deviations from the fixed point, $\delta y := y-z$, are determined by:
\begin{equation}
\label{mkg2}
\delta \dot{y}=\bar{f}(\delta y_{\tau_0}+z_0)-\frac{1}{T}\delta y.
\end{equation}
The fixed point $\delta y_0$ of (\ref{mkg2}) 
is given by $\delta y_0=T\bar{f}(\delta y_0+z_0)$. 
Therefore, the magnitude of the deviations from the fixed point $z_0$ decrease
for $T \rightarrow 0$ as shown in Fig. 3(a). According to eq. (\ref{mkg2}) the stability of 
the fixed point can be guaranteed for a small enough
$T$ as illustrated in Fig. 3(b). For $T \approx 2.0$, the fixed point $y_0$ undergoes a
Hopf bifurcation to an oscillating state.

In Fig. 4 we present the results of the successful control 
of the Mackey-Glass system from a low-dimensional chaotic state $(\tau_0=10)$ to a
chaotic solution of the R\"ossler system, $z(t)=x_1(t)$,
where $(\dot{x}_1=0.2x_1-x_2;\dot{x}_2=x_1-x_3;\dot{x}_3= \epsilon +4x_3(x_2-2))$.
In the case of a control towards a non-stationary goal dynamics $z(t)$, the quality of
the control is affected by the time scale 
of the goal dynamics. Loosely speaking, the parameters of the
control has to be adjusted such that the 
control is faster, than the a the fastest time scale of
the goal dynamics. This will be presented in more detail elsewhere.

\section{Outlook and Conclusions}

Let us shortly comment on the control of UPOs with a vanishing control force, 
which attracted most attention so far in nonlinear dynamics. 
The results of the paper will be also 
applicable in this case, if we choose the goal
dynamics to be: $z=z^{(T)}$, where $z^{(T)}$ 
is a periodic orbit of the system of period
$T$, which can be reconstructed from time series with well-known techniques \cite{UPO}. 
To state our result clearly, {\it all} UPOs of a scalar time-delay
system, even the ones which are embedded 
in a very high-dimensional chaotic attractor
are accessible to the control with a single control variable only. 
Furthermore, we would like to emphasize that the above arguments apply to {\it any}
solution of a time-delay system, indicating the possibility of controlling to non-stationary
and non-periodic (i.e. transients, high-dimensional chaos) trajectories with a vanishing control force.
The advantage of previously published methods for the control
of UPOs definitely is some sort of adaptive recognition of UPOs requiring
only a minimum of prior information about the system. In the framework of our control method, 
a control of UPOs of time-delay systems can be achieved by
requiring a suitably chosen functional upon the control force $F[z^{(T)}](t)$ 
to be minimal under the variation of the goal dynamics
$z^{(T)}$, i. e. ,
\begin{equation}
\label{UPO}
\dot{y}=H^{-1}(y,y_{\tau_0},\inf_{z^(T)} \int_{t-D}^t d \tau \| F[z^{(T)}](\tau) \|),
\end{equation}
where the functional is chosen to be an integration over the magnitude of the control force on an interval $[t,t-D]$. 
Technically, the minization can be performed 
using established nonlinear optimization 
procedures. 

To the authors knowledge there is no mathematical statement linking the number of {\em necessary} control variables to the invariant quantities of the controlled dynamics (such as attractor dimensions or number of positive Lyapunov exponents).
It is still an open question how many control variables are necessary
to ensure the unique existence of a control force, which
gains the control of hyperchaotic chaotic states. In some cases it has been shown numerically that a single control force is sufficient \cite{Pyragas,ses-single,tds-control}; in other cases there is evidence that more than one control force is required \cite{ses-multiple}. 
It seems to be a {\em sufficient} condition, as argued by several authors,
to apply a number of control variables equal to the number of positive Lyapunov-exponents. 
Besides that in practical purposes this turns out to be unhandy, we show in this paper that
the high-dimensional chaos of a $N$-component time-delay system can always be 
controlled with a $N$-component  control
force. Indeed, the number of positive Lyapunov-exponents is, in general,  not a meaningful 
quantity in this context. We believe that the number of "localized nolinearities" as introduced
in \cite{loc_nonlin}, which for time-delay systems corresponds to the number of components, 
might be a more appropriate concept to determine the number of control
variables.

Since time delays are common phenomena in nature, as well as in industrial processes,
we find a wide range of applications for the control of time-delay systems 
stemming from such different fields as 
hydrodynamics \cite{hydrodynamics}, laser physics \cite{laser}, physiology \cite{physiology}, 
and biology \cite{biology}. 
%We note in passing that the presented results do also apply for systems with
%several delay times, a continuous distribution of delay times, or certain
%spatially extended systems \cite{loc_nonlin}. 
For instance, we envisage the possibility to suppress the large-amplitude oscillations
of the high-Reynolds number turbulence observed in the Rayleigh-Benard system \cite{hydrodynamics}, 
with the time-varying temperature gradient as the control force. The same argument applies for
the control of a confined jet \cite{hydrodynamics} 
with the pressure difference as a control force.
Equally, the control of the high-dimensional 
chaotic states of a nonlinear ring resonator \cite{laser},
described by the Ikeda model, with the laser pump power as the control force,
seems to be possible with our control method.
Furthermore, in implementing the control in an adaptive manner,
a control for non-stationary time-delay systems is envisaged.

Finally, we summarize the conditons for the controllability of time-delay systems:
(1) To control a $N$-component time-delay system, it is sufficient to apply 
$N$ control forces.
(2) The function $H(y,y_{\tau_0},F)$ has at least to be locally invertible with respect to $F$
in the required range. If the underlying time-evolution of the delay system is not known, 
one has to rely on identification procedures, which additionally 
require: (1) The measurement of $N$ time series. (2) appropriate ranges for the identification.
The control is stable against a small amout of noise, 
as well as structurally stable against small deviations of
the model.

The author acknowledges useful discussions with W. Just, H. Kantz, A. Kittel, Th. Meyer,
J. Parisi, and A. Politi.

\newpage
\section*{Figure captions}

\begin{itemize}

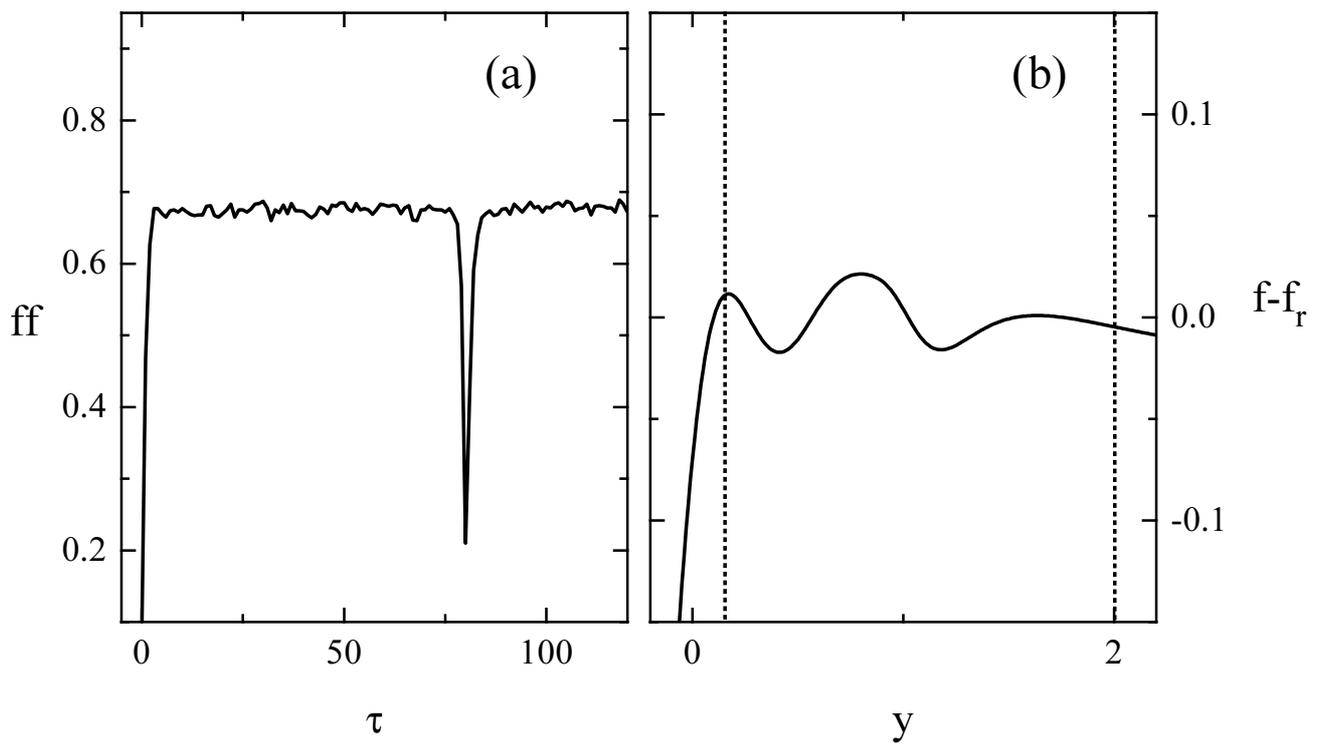
\item[{\bf Fig. 1:}] (a) Filling factor under variation of the delay time $\tau$ 
	for 1000 equally sized cubes in a three-dimensional space. 
	(b) The deviation of the fit $f_r$ from the function $f$. The dotted lines
	indicate the range ${\cal D}_1$.

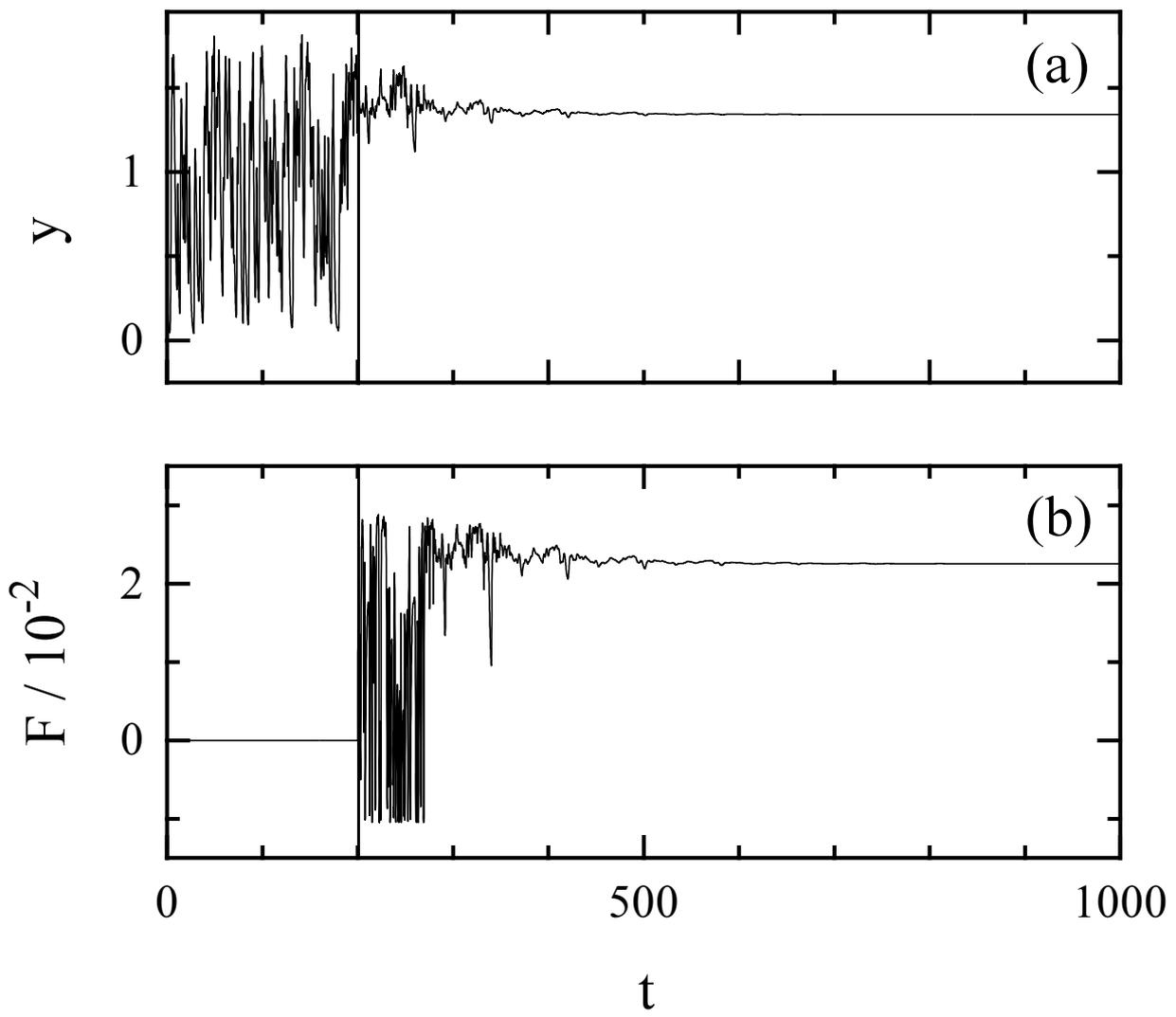
\item[{\bf Fig. 2:}] Control of the Mackey-Glass system for $\tau_0=80$ to the fixed point $z_0=1.5$. 
	Starting from a typical initial condition on the attractor ({\it transient time: 1000}), the 
	control has been switched on at $t=200$ (control parameter $T=1$)
	as indicated by vertical lines: (a) Time series $y(t)$; (b) Control force $F(t)$.
	
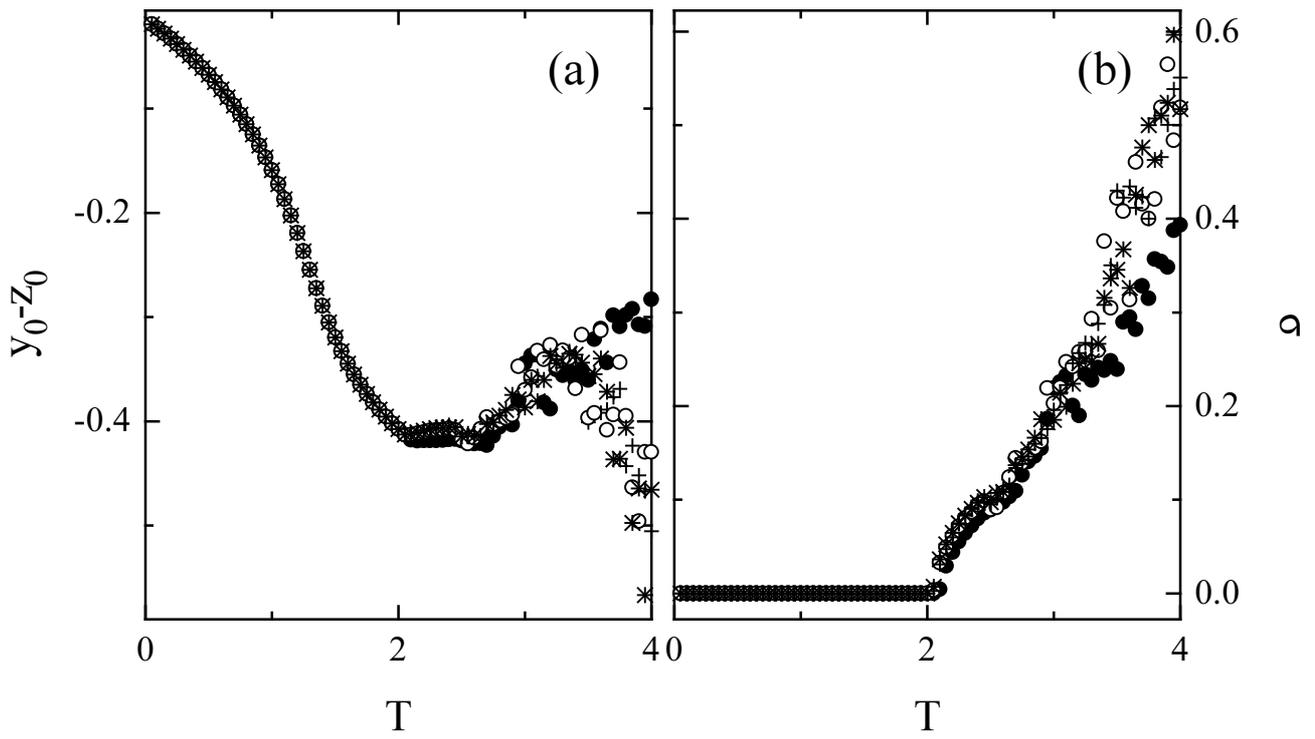
\item[{\bf Fig. 3:}] Control of the Mackey-Glass system under variation of 
	$T$ for $\tau=10$ ({\em solid circles}),
	$\tau=30$ ({\em open circles}),$\tau=50$ ({\em stars}), and $\tau=80$ ({\em crosses}): 
	(a) deviation of the fixed point $y_0$ from the goal dynamics $z_0$;
	(b) standard-deviation of the controlled dynamics.

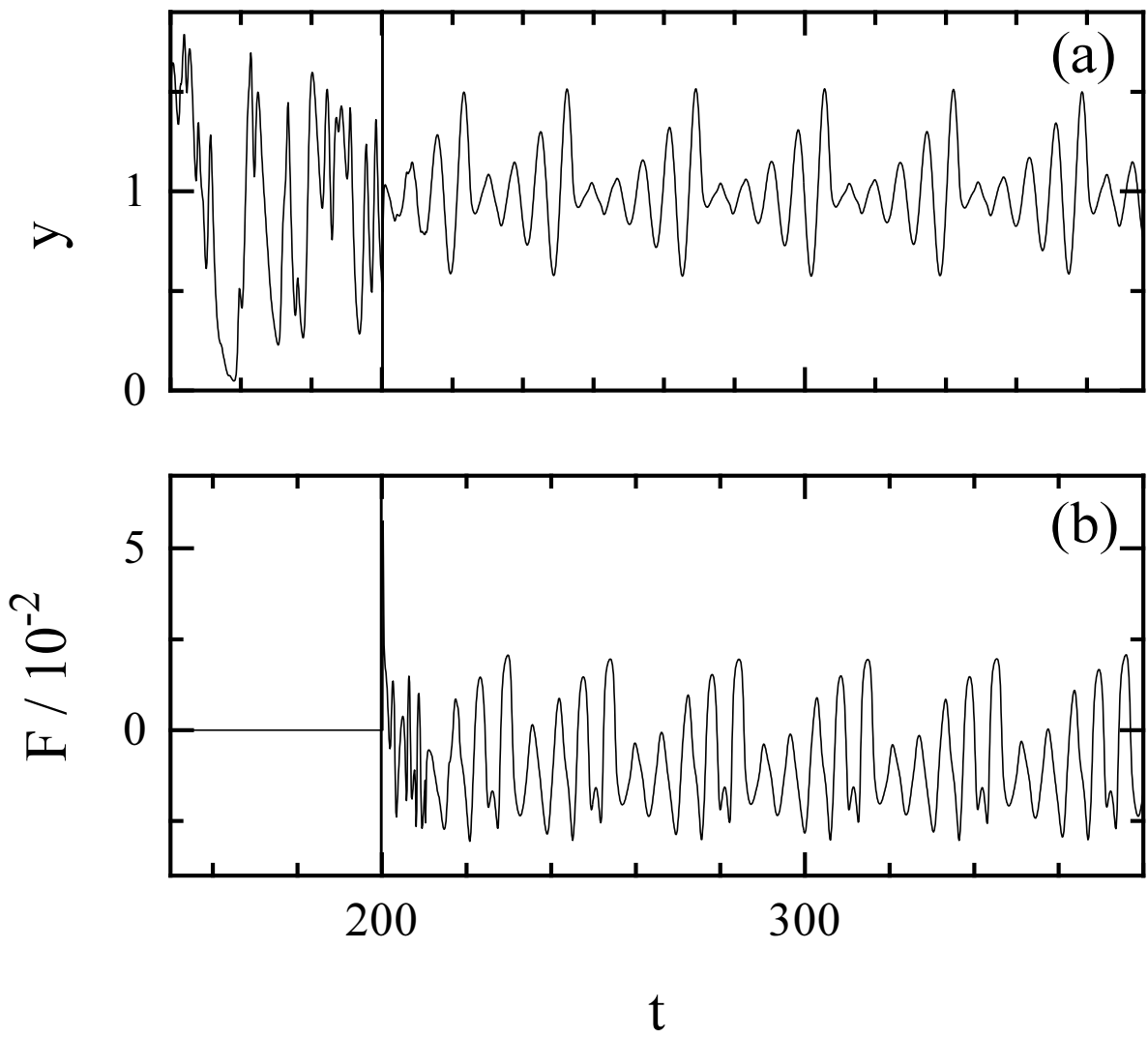
\item[{\bf Fig. 4:}] (a) Control of the Mackey-Glass system with $\tau_0=10$ to a
	variable of the R\"ossler system (not shown). The parameter of the control
	has been $T=0.10$. (b) Control force $F(t)$. As indicated by lines the control has been
	switched on at $t=200$.

\end{itemize}

\end{document}